\documentclass[5p]{elsarticle} 

\usepackage{lineno,hyperref}
\usepackage{footnote}
\modulolinenumbers[5]

\usepackage{scalefnt}
\usepackage{graphicx}
\usepackage{amsmath}
\usepackage{float}
\usepackage{xcolor}
\usepackage{array}  

\usepackage[utf8]{inputenc}

\journal{Results in Physics}
\begin{document}

\begin{frontmatter}

\title{The  Kardar-Parisi-Zhang exponents  for the $2+1$ dimensions}

\author[add1]{M\'arcio S. Gomes-Filho}
\ead{marcio@setubal.net.br}
\author[add1]{Andr\'e L. A.  Penna }
\ead{penna.andre@gmail.com }
\author[add3]{Fernando A. Oliveira}
\ead{fao@fis.unb.br}

\address[add1]{Instituto de F\'isica, Universidade de Bras\'ilia, Bras\'ilia-DF, Brazil.}

\address[add3]{Instituto de F\'{i}sica, Universidade Federal da Bahia, Campus Universit\'{a}rio da Federa\c{c}\~{a}o, Rua Bar\~{a}o de Jeremoabo s/n, 40170-115, Salvador-BA, Brazil.}

\begin{abstract}
The Kardar-Parisi-Zhang (KPZ) equation  has been connected to a large number of important stochastic processes in physics, chemistry and growth phenomena, ranging from classical to quantum physics. The central quest in this field is the search for ever more precise universal growth exponents. Notably, exact growth exponents are only known for $1+1$ dimensions. In this work, we present physical and geometric analytical methods that directly associate these exponents to the fractal dimension of the rough interface. Based on this, we determine the growth exponents for the $2+1$ dimensions,  which are in agreement with the results of thin films experiments and precise simulations.  We also make a first step towards a solution in $d+1$ dimensions, where our results suggest the inexistence of an upper critical dimension.
\end{abstract}


\end{frontmatter}

\section{Introduction}

In most physical systems, the growth process occurs when particles, or aggregates of particles, reach a surface via diffusion, an injection beam or some kind of deposition process. To investigate the growth, we follow the height  $h(\vec{x},t)$, where $t$ is the time and  is the position in a space of dimension d. Since  $h(\vec{x},t)$ has scaling properties different from  $\vec{x}$,  we say that $(h(\vec{x},t),\vec{x})$ forms a $d+1$ dimensional space. Field equations have been proposed for the  $h(\vec{x},t)$ dynamics such as the Kardar-Parisi-Zhang (KPZ) equation~\cite{Kardar86}:
\begin{equation}
\label{KPZ}
\dfrac{\partial h(\vec{x},t)}{\partial t}=\nu \nabla^2 h(\vec{x},t) +\dfrac{\lambda}{2}[\vec{\nabla}h(\vec{x},t)]^2+ \eta(\vec{x},t),
\end{equation}
where the  Gaussian white noise, $\eta(\vec{x},t)$,  has zero mean $\langle \eta(\vec{x},t) \rangle = 0$ and variance
\begin{equation}
\label{FDT}
\langle \eta(\vec{x},t) \eta(\vec{x'},t') \rangle  = 2D\delta^{(d)}(\vec{x}-\vec{x'})\delta(t-t').
\end{equation}
The coupling parameter $g=D\lambda^2/\nu^3$ connects the KPZ coefficients, being $\nu$ (surface tension) associated with the Laplacian smoothing mechanism and $\lambda$ related to the tilt mechanism.

A large number of phenomena~\cite{Barabasi95,Merikoski03,Doussal16,Orrillo17,Ojeda00,Chen16} can be understood by defining a few physical quantities such as the average height $\langle h \rangle$ and  the roughness or surface width
\begin{equation}
\label{W2}
 w(l,t)^2= \langle h^2(t)\rangle -\langle h(t) \rangle^2,
\end{equation}
where $l$ is the sample size. We are interested in physical systems in which the roughness grows with time and afterwards saturated with a maximum value $w_s$, i.e.~\cite{Barabasi95}:
\begin{equation}
\label{Sc1}
w(l,t)=
\begin{cases}
 ct^\beta , &\text{ if~~ } t \ll t_\times\\
 w_s \propto l^\alpha, &\text{ if~~ } t \gg t_\times,\\
\end{cases}
\end{equation}
being $t_\times \propto l^z$ and the exponents  related by:
\begin{equation}
\label{z}
z=\frac{\alpha}{\beta},
\end{equation}
and by Galilean invariance~\cite{Kardar86}:
\begin{equation}
\label{GI}
\alpha+z=2.
\end{equation}

The KPZ approach describes and connects a wide range of experiments~\cite{Merikoski03,Orrillo17,Ojeda00,Chen16,Takeuchi13,Almeida14, Almeida17, Fusco16} and models \citep{Oliveira13,Calabrese10,Amir11,Rodriguez19, Mello01,Rodrigues14,Rodrigues15,Reis04,Alves16,Gomes19,Meakin86,Krug92,Krug97,Derrida98,Daryaei20}.
 Note that most of these processes are interconnected, as an example, the single step (SS)  model~\cite{Meakin86,Krug92,Krug97,Derrida98,Daryaei20} is connected with
the asymmetric simple exclusion process~\cite{Derrida98}, the six-vertex model~\cite{Meakin86,Gwa92}, and the kinetic Ising model~\cite{Meakin86,Plischke87}. Recently, quantum versions of KPZ equation have been formulated~\cite{Corwin18,Ljubotina19, DeNardis19,Nahum17}.  Despite all effort, we are still far from a satisfactory theory for the KPZ equation, what makes  it one of the  toughest  problems in modern mathematical physics, and probably one of the most important problem in nonequilibrium statistical physics, for reviews see~\cite{Krug97,Derrida98,Meakin93,Oliveira19,Healy15,Henkel17}.
Thus,  two main questions are still open:\\
    \hspace*{0.5cm}$1.$ What is the probability of height distributions?\\
    \hspace*{0.5cm}$2.$ What are the exponents $\alpha$, $\beta$ and $z$?\\
Up to now  these questions have been answered exactly only in $1+1$ dimensions. The first due to the succession of researches~\cite{Doussal16,Calabrese10,Amir11,Prahofer00,Sasamoto10} and the second by the original KPZ work~\cite{Kardar86}.  In this  work we propose to obtain the exponents for a $2+1$ dimensional system from first principle, precise simulations and analysis of data in the literature.

\section{Theory}
Let us start considering that  if we know one of the exponents $(\alpha,\beta,z)$, we can determine the others from the relations (\ref{z}) and (\ref{GI}), in a such way that we will concentrate  in the determination of the roughness exponent $\alpha$.
 In the search for the KPZ exponents several analytical methods, such as scaling relation and renormalization group (RG) approaches have been tried. Up to now we can resume that as:
\begin{enumerate}
 \item Scaling fails for all dimensions.
 \item RG works only for $1+1$ dimensions~\cite{Kardar86}. It fails for all $d>1$.
  \item Field theoretical methods yield exponents that are not precise~\cite{Lassig98}.
\end{enumerate}
  By the ``failure'' of these approaches, we mean that they  were not able to give precise exponents. However, some calculations using   RG produces very useful results. For example, Canet {\it et al}~\cite{Canet10,Canet11} used nonperturbative RG and obtained the only complete analytical approach yielding a qualitatively correct phase/flow diagram to date. We shall consider what is particularly valid in these approaches and we shall add three important ingredients: first, we use dimensional analysis which is always stronger, fractal dimension of the interface and  a proper correction of the fluctuation dissipation theorem for higher dimensions.

In this way, we consider here the full space dimension as $\tilde{D}=d+1 $ and the fractality of the   interface, which has a total dimension $\tilde{D}_i=d_f+1$,  where $d_f$ is a fractal dimension~\citep{Barabasi95}. It is well known that the fractal dimension is related to the exponent $\alpha$ by~\cite{Barabasi95,Kondev00}:
\begin{equation}
\label{df2}
\alpha=2-d_f,
\end{equation}
for $d=1,2$. 
Note that we just need a second equation relating  $d_f$ to $\alpha$ to obtain both. Now, we are going to consider the saturated roughness:
\begin{equation}
\label{EW}
w_s=\sqrt{\frac{D}{24 \nu} l},
\end{equation}
which was obtained from  the seminal work of  Krug {\it et al.}~\cite{Krug92} in  the  universal behavior  of driven interfaces  in $1+1$ dimensions. This result is exact for $1+1$ dimensions and it is our starting point for our analysis in $d+1$ dimensions.

\subsection{Dimensional analysis}

 Dimensional analysis is a powerful technique, we will apply it here combined with scaling and the concept of universality.
First note  $l$ must be associated with a physical size, then we shall take it as  $l  \rightarrow al$, where $a$ is lattice space with dimension $[a]=[L]$, being $[L]$ the space dimension and $l$ just a number, i.e. the number of lattice spaces. Hence we  generalize the  Eq.~(\ref{EW}) for  KPZ in $d+1$ dimensions as:
\begin{equation}
\label{wsg}
w_s= \left( \frac{D}{24\nu}a l \Psi \right)^\alpha,
\end{equation}
where $\Psi$ is a dimensionless number. Nevertheless, dimensional analysis does not give us the nondimensional quantities, it is convenient to keep Eq.~(\ref{wsg}) in the above form. First, we shall distinguish dimensional analysis from scaling, for example  $w_s$ scales as $l^\alpha$, however its physical dimension is the same as the height $h$,  $[w_s]=[h]=[L]$, i.e in experiments they are measured in units of length, as it must be from definition (\ref{W2}). For example, in experiments of  cadmium telluride thin growing on Si~$\langle 001 \rangle$ surfaces~\cite{Almeida14}. While the substrates have size $10 \mu m  \times 10 \mu m $, the  $h \sim nm$ and  $w \sim nm$, which means that both quantities scale different, but their physical unit remains in length units.

\subsection*{1+1 dimensions}  In order  to prove the above relation,  let us first perform  the  dimensional analysis in $1+1$ dimensions. The physical dimensions involved are  $[\nu]=[L^2][T^{-1}]$, $[D]=[L^{3}][T^{-1}]$ and $[\lambda]=[L][T^{-1}]$ where $[T]$ is the time dimension. Considering  all KPZ parameters, we rewrite Eq.~(\ref{wsg})  as:
\begin{equation}
\label{werr}
 w_s=\left[ \frac{al}{24}D^{\phi_1 } \lambda^{\phi_2} \nu^{\phi_3} \right]^\alpha=\left[ \frac{al}{12}A \Psi(\lambda) \right]^\alpha,
\end{equation}
where
\begin{equation}
\label{A}
      A=\frac{D}{2\nu}.
\end{equation}
Since $w_s$ must be time independent,  we have  $\phi_3= -(\phi_1+\phi_2)$ and
\begin{equation}
 \Psi(\lambda)= \left ( \frac{D}{\nu} \right )^{\phi_1-1} \left ( \frac{\lambda}{\nu} \right )^{\phi_2}.
\end{equation}
Thus, we get:
\begin{equation}
\label{alf0}
\alpha=\frac{1}{2+\phi},
\end{equation}
with   $\phi=(\phi_1-1)-\phi_2$.   
Note that in order to have Eq.~(\ref{werr}) equal to Eq.~(\ref{wsg}) we must have  $\Psi=1$, which gives $\phi_1=1$ and $\phi_2=0$, which implies in $\phi=0$, and one recovers the Krugs result (\ref{EW}) with $\alpha=1/2$.

\subsection*{d+1 dimensions}

Before we continue our analysis, let us remember that the fluctuation relation Eq.~(\ref{FDT}) works for $d=1$~\cite{Kardar86,Rodriguez19,GomesFilho21}, however it does not work properly for higher dimensions. The violation of the FDT is well-known in the literature, for example, in  KPZ~\cite{Kardar86,Rodriguez19, Anjos21} for $d>1$, in structural glass~\cite{Grigera99} and in ballistic diffusion~\cite{Costa03,Vainstein06,Costa06,Lapas07,Lapas08}. Note that all parameters $w_s$, $a$ and $\nu$ have a well fixed time and space dimension, the only one that change with the space dimension $d$ is the fluctuating or noise parameter $D$ and, therefore, we expect some violation of fluctuations relations, such as~(\ref{FDT}).

In growth we may have a simple answer for that, consider for example the SS model, which  is defined in such way that the height difference between two neighboring heights $\eta=h_i-h_j$ is just $\eta=\pm 1$.  Now, let us consider a hypercube of side $L$ and volume $V=L^d$. Thus, we can select a site $i$ and compare its height with that of its neighbors $j$, applying the following rules:
\begin{enumerate}
\item At time $t$, randomly choose a site $i\in{V}$;
\item If $h_i(t)$ is a minimum, then $h_i(t+\Delta t)=h_i(t)+2$, with probability $p$;
\item If $h_i(t)$ is a maximum, then $h_i(t+\Delta t)=h_i(t)-2$, with probability $q$.
\end{enumerate}

Note that in $(1)$ we have a white noise in the $d+1$ space. However, due to rules $(2)$ and $(3)$ only a part of the noise will be  really  effective. An analogy is to throw a beam of light on a surface. For a flat surface we get the same beam reflected. However, if the surface is rough, the reflected beam will be completely modified. Thus, we must consider not the applied noise, but the noise selected by the rough surface (the system's response), which can have different properties, such as intensity and dimension.

\begin{figure}[htbp]
\centering
\includegraphics[width=1\columnwidth]{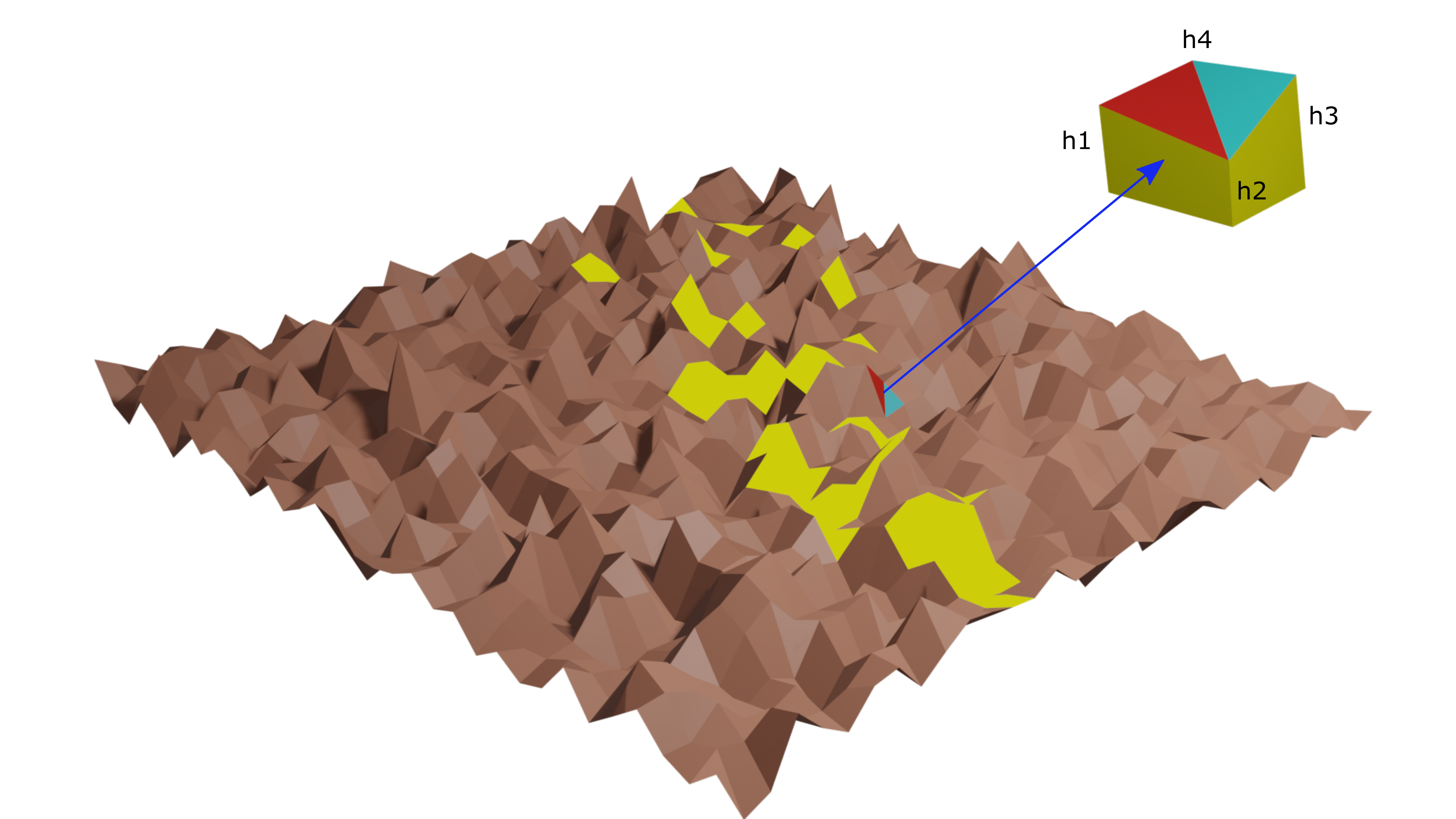}
\caption{The fractal geometry of the surface for $2+1$ dimensions.  The  unit cell used to compute the fractal dimension is highlighted, being $h_1=h(x,y)$,  $h_2=h(x,y+\Delta y)$, $h_3=h(x+\Delta x,y+\Delta y)$ and $h_4=h(x+\Delta x,y)$. For a fractal geometry, the total size of the surface $S_T$ is given by $S_T \propto \Delta l^{d_f-d}$~\cite{Barabasi95}, here $\Delta x = \Delta y= \Delta l$.} 
\label{figfractal}
\end{figure}

The second and more strong reason to change the dimension of the response is the widely known fact that the surface has a fractal dimension~\cite{Barabasi95,Kondev00}, as already mentioned above, and as exhibit  for example in  SiO$_2$ films~\cite{Ojeda00} or in the rough interface generated for the $2+1$ SS  model~\cite{Daryaei20}. 
To illustrate that,  we show in Fig.~(\ref{figfractal}) the fractal geometry for $2+1$ dimensions and the method described to obtain the fractal dimension.

Now, under this condition the noise intensity at the interface will have the dimension $[D]=[L^{df+1}][T^{-1}]$.
Thus,  the dimensional analysis yields now:
\begin{equation}
\label{alf1}
\alpha=\frac{1}{d_f+1+\phi},
\end{equation}
with
\begin{equation}
 \phi=(\phi_1-1)d_f-\phi_2.
\end{equation}

\subsection*{Universality}
 For $d+1$ dimensions, as the parameters $\nu$ and $\lambda$  are not universal, i.e. they change with the model, in the particular case of the SS model, they are function of the probability $p$.  If each model has different values of $\phi$ that results in different values of $\alpha$, which contradicts  the KPZ universality. Thus  all models in the KPZ universality classe must have $\phi=0$.   On the other hand, there is a less restrictive solution with $\Psi \neq 1$ that preserves universality, i.e. $\phi=0$, with
\begin{equation}
\label{phi}
 \phi_1=1+\phi_2/d_f,
\end{equation}
for that we do not need to know what the exponent $\phi_2$ is, and the parameter $\Psi$ is just a number, which may change from model to model. Consequently, without contradicting universality  $\Psi \neq 1$ is not out of the cards. 
 Therefore, it follows that the dimensional analysis with universality, $\phi = 0$, determines the roughness exponent as:
 \begin{equation}
\label{alf}
\alpha=
\begin{cases}
 1/2 , &\text{ if~~ } d=1\\
\frac{1}{d_f+1}, &\text{ if~~ } d \geq 2.\\
\end{cases}
\end{equation}


 Thus a RG  approach for $d+1$ dimensions must consider the fractality of the interface\footnote{Although it is not our objective here to do a full RG for KPZ, we have done a first draft of this RG approach with a fractal noise, within one loop expansion. The integrals becomes more complicate, but the contributions due to corrections in $\lambda$  sums up zero as in the KPZ classical work~\cite{Kardar86}. Thus, the Galilean invariance~(\ref{GI}) is still valid.}.
 
 However, for $d > 2$, the available simulation data in the literature for the exponents are very rough, and we shall focus our analysis in $2+1$ dimensions.

\subsection*{2+1 dimensions}

 The $2+1$ dimensions is  the most  important one, besides being our real world, the growth phenomena are associated to surface science, also to the development of new technological devices, such as thin films. There are experiments and more simulation results available and we can get more precise exponents than for $3 + 1$, for example. For the  EW  equation the fractal dimension is $d_f=2$, in this case from Eq.~(\ref{df2}) we get $\alpha=0$, in agreement with the scaling relation $\alpha=(2-d)/2=0$~\cite{Barabasi95}, but in strong disagreement with the relation~(\ref{alf}).  However, a null exponent in phase transition and also in diffusive process~\cite{Oliveira19} does not mean a $w_s$  independent of $l$, but rather a logarithmic  behavior, $w_s \propto \ln{(l)}$, being this behavior  recently observed for EW model~\cite{Daryaei20}.  In this case, the power law behavior suggested in~(\ref{wsg}) is not valid, and we are only concerned here with situations in which the power law behavior holds.

 Now we return to the KPZ equation, where the equations (\ref{alfn}) and (\ref{df2}) for $d=2$ yield:
\begin{equation}
\label{exp}
 \alpha=\frac{3-\sqrt{5}}{2}; \hspace{0.5cm}  \beta=\sqrt{5}-2;  \hspace{0.5cm}z =d_f=\frac{1+\sqrt{5}}{2},  
\end{equation}
which gives   $d_f=1.61803...$ and $\alpha=0.381966011...$. The  exponents above are our major results.

\section{Simulations Results}

\begin{figure}[htbp]
\centering
\includegraphics[width=1\columnwidth]{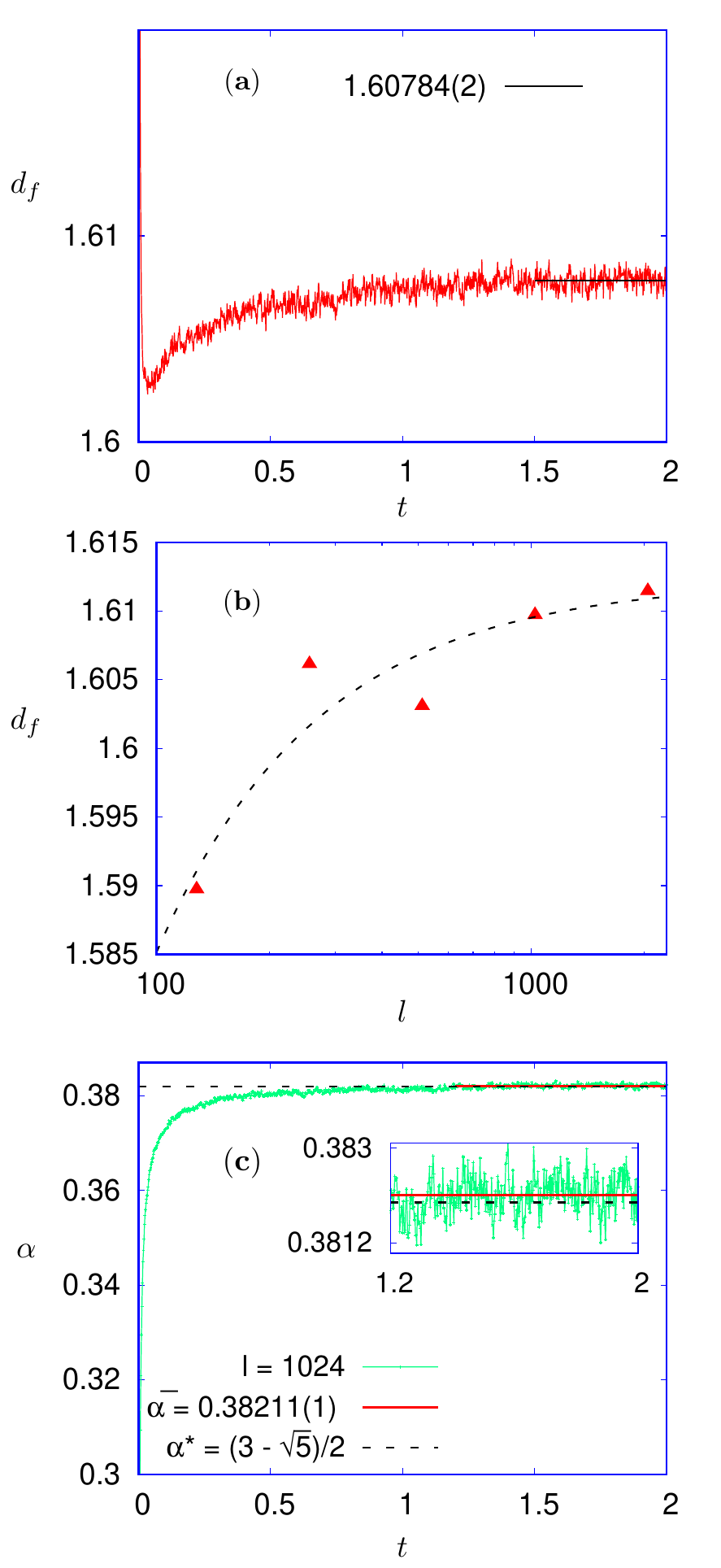}
\caption{ Etching model in $2+1$ dimensions: $(a)$ $d_f$ against time $t$ (in units of $t_\times$). $(b)$ the semi-log plot of the $d_f$ in function of the size $l$. The adjusted curve was obtained from the fit of the function $d_f(l)=d_f-c/l$. (c) $\alpha$ as a function of time $t$ (in units of $t_\times$)  obtained from the correlation function~(\ref{Corr}), being  the dashed line the analytical result, Eq.~(\ref{exp}).}
\label{figdimfrac}
\end{figure}

In order to have some precise numerical results, let us consider the etching model~\cite{Mello01,Rodrigues14,Alves16} for $2+1$ dimensions. This model belongs to the KPZ universality class~\cite{Gomes19}. First, we obtain the fractal dimension, $d_f$, from the boxing counting method~\cite{Barabasi95}, as it is depicted in Fig.~(\ref{figfractal}). The results were averaged over the number of sites and also over different numerical experiments. The number of experiments is given by   $N_e=(2^{9}/l)^{5/2}$, and a single experiment for $l \ge 2^9$, in such way that a large sample needs a small number of experiments.

We show in Fig~\ref{figdimfrac}~$(a)$ $d_f$ as function of the time  for a square lattice of size $l=2^{11}$.  After convergence, we get the time average, which is represented by the vertical black line, to obtain  $d(l,t \rightarrow \infty)= d_f \equiv d_f(l)$,   for each value of $l$.  In $(b)$, the semi-log plot of the $d_f$ as function of the size $l$. In order to correct the  finite size effects,  we adjust the points to $d_f(l)=d_f-c/l$ and obtain $d_f=1.612(2)$, which inserted into Eq.~(\ref{alf}) yields  $\alpha=0.3828(3)$.
It is important to note that these values are very close to the analytical results  $d_f=\frac{1+\sqrt{5}}{2}=1.61803$ and $\alpha=0.381966011..$.   Already in Fig.~\ref{figdimfrac}~$(c)$, we exhibit $\alpha$ versus time $t$  for a square lattice of size $l = 2^{10}$ obtained from the correlation function:
\begin{equation}
\label{Corr}
C(r,t)= \left\langle [h(\vec{x}+\vec{r},t)- h(\vec{x},t)]^2\right\rangle \propto r^{2\alpha},
\end{equation}
where $r$ is the modulus of the vector $\vec{r}$ with $r < \xi$, being $\xi$ the correlation length~\cite{Kondev00}. In this sense, we consider the first $9$ neighbors along the principal axes, the results were  averaged  over $1000$ different simulations and $\alpha(t)$ was found by fitting the correlation curve. $\alpha$ increases rapidly with time and afterwards equilibrated with its values fluctuating around the mean, $\overline{\alpha}= 0.38211(1)$, where it was estimated in the range of $1.2 \leq t \leq 2$ (solid line),  being this value very close to our analytical values (dashed line), as it is depicted in the inset of Fig.~\ref{figdimfrac}~$(c)$.

\begin{figure}[htbp]
 \centering
 \includegraphics[width=1\columnwidth]{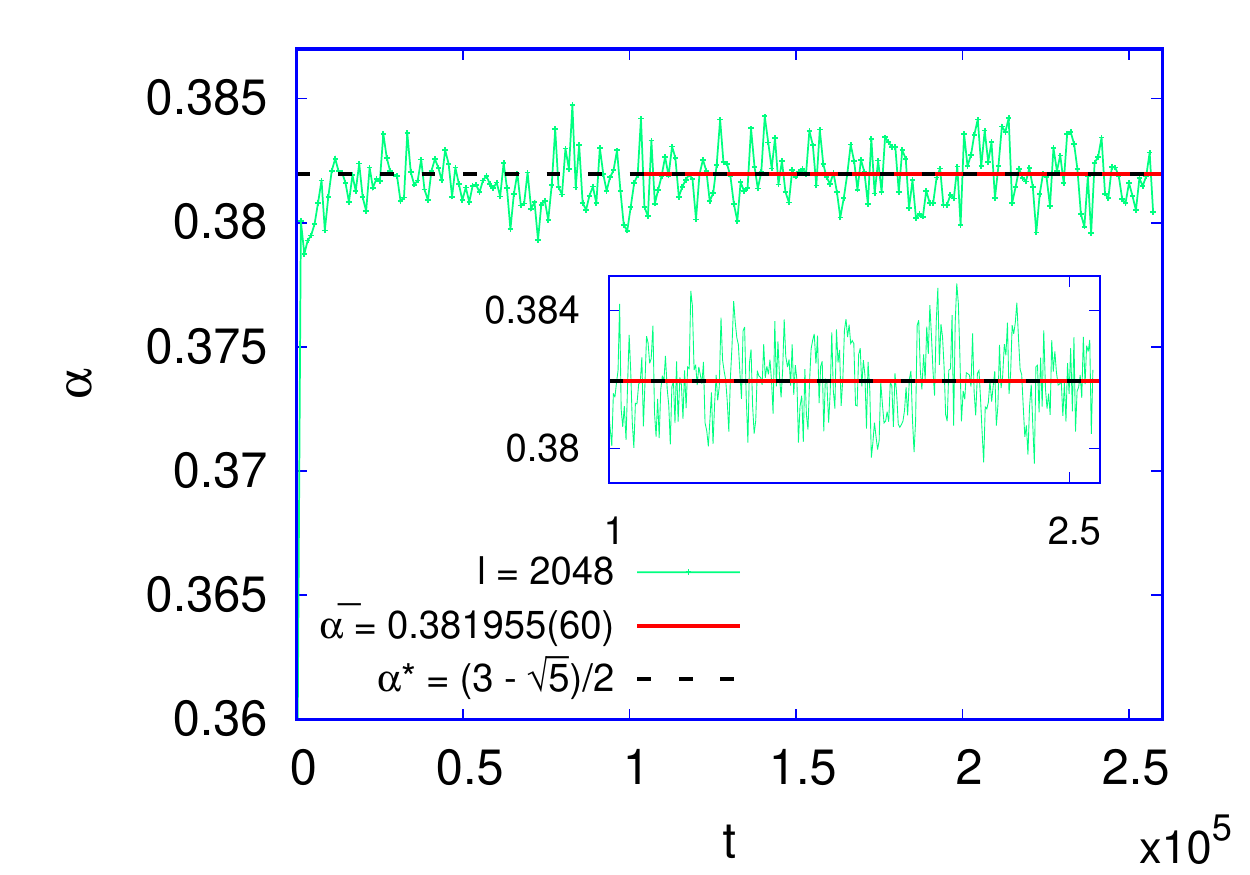}
 \caption {\label{fig:alphaSS} SS model: the exponent $\alpha$ as a function of time  $t$   for a square lattice of size $l = 2^{11}$ obtained from the correlation function~(\ref{Corr}) where we consider the first $3$ neighbors along the principal axes. The simulation time here is expressed in units of deposition layers.}
\end{figure}

A remarkable result was found for the SS model for a system with $p=1$ and size $l=2048$ in which the results were averaged  over the number of sites, $3$ experiments and again over time to obtain $\overline{\alpha}=0.381955(60)$, as shown in  Fig.~\ref{fig:alphaSS}.

 \begin{figure}[htbp]
  \centering
  \includegraphics[width=1\columnwidth]{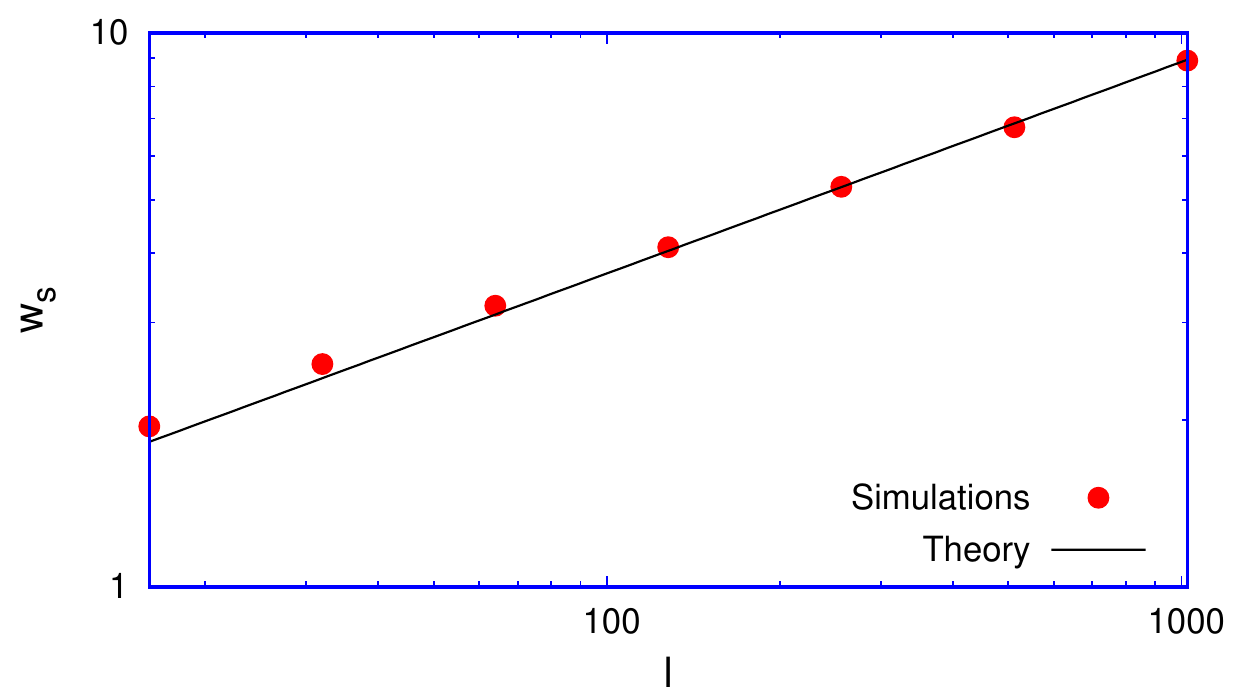}
  \caption { \label{fig:ws} Log-log plot of $w_s$ as function of $l$ for $2+1$ etching model (points) and the line was obtained from Eq.~(\ref{wsg}), being  $A=3.629(9)$~\cite{Carrasco14} and  $\alpha=0.381966011..$, Eq.~(\ref{exp}). For small $l$ we see some finite size effects.}
\end{figure}

In addition, in Fig.~(\ref{fig:ws}) we show $w_s$ versus $l$ for the $2+1$ etching model.  We use the value $A=3.629(9)$  from~\cite{Carrasco14} to obtain  $\alpha=0.3815(2)$.  From the etching model~\cite{Gomes19} using $\frac{D}{2\nu}=3.62(3)$ thus Eq.~(\ref{A}) yields $\Psi=1.00(2)$. It should be noted that this result confirms not only the exponent but also the factors within Eq.~(\ref{werr}), the second part, $\Psi=1$, up to now only  for the etching model. However, the major point here is that the factor $\Psi$ does not alter our dimensional analysis, therefore the exponents are independent of it.

We have found the simulations results for different models (etching and SS models) in a good agreement with the analytical results. In the following section we shall compare our analytical results with experimental/computational results from the literature.

\section{Literatura data}

\begin{table}[h!]
\scalefont{0.8}
 \begin{tabular}{c c c c}
\hline
\hline
Authors		        &   $\alpha$               & $\delta$          & Model \\
PW(A)               &   $\frac{3-\sqrt{5}}{2}$ & $0$               & Universal \\
PW(b)               &    $0.3828(3)$           &  $-0.008(3)$      & Etching  \\
PW(c)               &    $0.38211(1)$         &  $-0.00014(1)$    & Etching   \\
PW(d)               &    $0.3815(2)$           &  $0.005(2)$       & Etching  \\
PW(e)               &    $0.381955(60)$           &  $0.000011(60)$       & SS  \\
 \cite{Oliveira13}  &   $0.387(4)$             &  $-0.005(4)$      & RSOS \\
 \cite{Oliveira13}  &   $0.386(6)$             &  $-0.004(6)$      & SS \\
\cite{Oliveira13}   &   $0.381(7)$             &  $0.001(7)$       & Etching \\
 \cite{Oliveira13}  &   $0.387(13)$            &  $-0.005(13)$     & RSOSC \\
 \cite{Oliveira13}  &   $0.374(26)$            &  $0.008(26)$      & SSC \\
\cite{Oliveira13}   &   $0.379(9)$             &  $0.003(9)$       & Eden (001)\\
\cite{Oliveira13}   &   $0.386(8)$             &  $-0.004(8)$      & Eden (111)\\
\cite{Reis04}     &   $0.383(8)$             &  $-0.001(8)$      & Etching\\
\cite{Kondev00}     &   $0.38(8)$              &  $-0.002(8)$      & RSOS \\
\cite{Pagnani15}     &   $0.3869(4)$            &  $-0.0049(4)$     & RSOS \\
 \cite{Kelling18}   &   $0.3889(3)$            &  $-0.0069(3)$     & Octahedron\\
\cite{Odor09}       &   $0.377(15)$            &  $0.005(15)$      &  DLC\\
\cite{Healy12}      &   $0.388(-)$             &  $-0.006(-)$      &  KPZ\\
\cite{Healy12}      &   $0.385(4)$             &  $-0.003(4)$      & DPRM \\
\cite{ChenShan99}   &   $0.38(1)$              &  $-0.002(1)$      & BCSOS \\
\cite{Colaiori01}   &   $0.38(-)$              &  $-0.002(-)$      & Mode-coupling-KPZ \\
\hline
\hline
\end{tabular}
 \caption{\label{Tab01}
 Values of $\alpha$ and deviation $\delta$. Here $\delta=\alpha^*-\alpha$, measures the deviation from the analytical value $\alpha*=\frac{3-\sqrt{5}}{2}$. PW(A) stands for Present Work Analytical results. PW(b) by determination of the fractal dimension, Fig.~\ref{figdimfrac}(b); PW(c) using the correlation function, Eq.~(\ref{Corr}), Fig.~\ref{figdimfrac}(c); PW(d) from Fig.~(\ref{fig:ws}), and PW(e) from the SS model,~Fig.~\ref{fig:alphaSS}.   Observe that the same method~\cite{Oliveira13}  can yield slightly different results, mainly fluctuations, for different models.  The average of all values gives $\overline{\alpha}=0.3828 \pm 0.0037$ and  $\overline{\delta}=-0.0008 \pm 0.0037$.}
\end{table}


\subsection*{Experimental results}  We shall analyze carefully the existing experimental results. First, on measuring $\alpha$, or computing it using~(\ref{Corr}) we must know in what regime we are. For example, one  should always remember that in the growth phenomena each time unit  corresponds to a deposition layer. Consequently, thin film will not  achieve the saturation regime of  Fig.~(\ref{fig:alphaSS}), where $t_x \sim 10^5$ layers. In addition, we must distinguish between  local and global value of $\alpha$~\cite{Edwin19}. Thus, a good exponent is obtained in experiments that measure the exponent $z$. Accurate experiments give  $z=1.6(2)$~\cite{Orrillo17}, $z=1.6(1)$~\cite{Ojeda00},   $z=1.61(5)$~\cite{Almeida14}, and $z=1.61$ \cite{Fusco16} in agreement with our value of $z =d_f=\frac{1+\sqrt{5}}{2}=1.61803...$. As the final destination of any theory is decided by the experiments,  we can say that so far the odds favor us.

\subsection*{Computational and theoretical results }

Finally, in order to provide how much our analytical value, $\alpha^*=(3-\sqrt{5})/2=0.381966011...$, is  close to
 $\alpha$ values for some models in $2+1$ dimensions, as listed in Table~\ref{Tab01}, we measure $\delta=\alpha^*-\alpha$, which is the deviation from the analytical value. Thus, as we can see, all of these values are close to our analytical results and $\delta$ oscillates around zero.

\section{d +1 dimensions}

Now we propose an extension of the Eq. (\ref{df3}) for $d \geq 2$ as
\begin{equation}
\label{df3}
\alpha=\tilde{D}-\tilde{D}_i=d-d_f.
\end{equation}
 Since $0<\alpha<1$, then $d-1<d_f <d$. This is what we expect for the dimension of one fractal embedded in an integer space.

Therefore, for $d+1$ dimensions with $d \geq  2$, the Eq.~(\ref{df3})  together with the Eq.~(\ref{alf}) yields:
\begin{equation}
\label{alfn}
d_f =  \frac{d-1+\sqrt{\Delta}}{2}, \hspace{0.7cm}   \alpha  =  \frac{d+1-\sqrt{\Delta}}{2}
\end{equation}
and
\begin{equation}
\label{bez}
z =  \frac{3-d+\sqrt{\Delta}}{2}, \hspace{0.7cm}   \beta  =  \frac{d-\sqrt{\Delta}}{3-2d}
\end{equation}
with $\Delta=(d+1)^2-4$. This simple relation give us the fractal dimension and  the exponents $(\alpha,\beta,z)$ for all $d \geq  2$. This shows as well that there is no upper critical limit for the KPZ equation, since the exponents decay with $d$ without showing any specific upper critical dimension.  Unfortunately, for $d \geq  2$, the available simulations data is not as precise as that of $2+1$.

\section{Conclusion}
 In this work, our objective was to obtain the most accurate  values of the growth exponents in 2 + 1 dimensions for the KPZ equation. In order to do that, we extend the Krug's universal solution for  $1+1$ dimensions  to $d+1$ dimensions  and  we impose the fractality of the interface,  by replacing the integer dimension $d$ by the fractal dimension $d_f$ of the roughness interface, from which  we were able to determine  the values of $2+1$ Kardar-Parisi-Zhang exponents. The solution provides the golden ratio $d_f=\frac{1+\sqrt{5}}{2}$, which also appears in many natural phenomena, such as growth of vegetable and animals. Moreover, we show that our results are not only in good agreement with our numerical results, but also with the literature data for some models in $2+1$ dimensions, which shows that the KPZ universality class holds.  For $2+1$ dimensions we believe that our proposal is fully justified. Considering that no approximation was done to obtain the exponents~(\ref{exp}), we believe that these are good candidates to be the exact values.   However,  for $d+1$ dimensions, with $d>2$ there is a lack of reliable results, thus new theoretical and simulations results will be necessary for higher dimensions.  Nevertheless, we have  achieved important progress.  First, we have explicit analytical results well confirmed by simulations for $2+1$ dimensions.  Better, it is in accordance with experimental results; second, notice that the exponents appear as irrational numbers and not as the ratio between integers as in the first guess~\cite{Wolf87,Kim89} and they are now directly related to the fractal dimension $d_f$. This is more close to what has been obtained from numerical simulations. Finally, the discussions presented here open a new scenario for further investigation of different forms of growth both theoretical and numerical.  For example, the RG approach in the fractal interface will probably originated new important results, as  mentioned above, one loop expansion preserves the Galilean invariance (\ref{GI}). However, it deserves further developments. The  attempt to obtain exact height fluctuations for the stationary KPZ equations, as well as for most of physics in  $2+1$ dimensions  is still in its  begin.  These theoretical methods will benefit  from the fixed points obtained by precise KPZ exponents, and from a fractal geometry that must be associated with them~\cite{Anjos21}. We expect as well that new methods would confirm our results. This work opens a new horizon for KPZ research.

\section*{Acknowledgments}

The authors are grateful for the many useful communications and advice from Profs. Janos Kertesz, Thiago A. Assis, Tiago J. Oliveira, and Francisco Alcaraz. We would also like to thank Mrs. Rayra S. S. Veloso for her kind review. This work was supported by the Conselho Nacional de Desenvolvimento Cient\'{i}fico e Tecnol\'{o}gico (CNPq), Grant No.  CNPq-312497/2018-0 and the Funda\c{c}\~ao de Apoio a Pesquisa do Distrito Federal (FAPDF), Grant No. FAPDF- 00193-00000120/2019-79.

\bibliography{references}

\end{document}